\documentstyle[aps,prl,twocolumn]{revtex}
\input psfig.tex
\begin{document}
\title{
Spectral Dependence of Polarized Radiation due to Spatial Correlations
}
\author{Krishna Bhuvalka and Pankaj Jain}
\vskip 0.5cm
\address{\it Department of Physics, IIT Kanpur, Kanpur-208 016, INDIA\\
}
\maketitle

\begin{abstract}
{\large\bf Abstract:}
We study the polarization of light 
emitted by spatially correlated
sources. We show that in general polarization acquires  
nontrivial spectral dependence due to spatial correlations.
The spectral dependence is found to be absent 
only for a special class of sources where the correlation length scales
as the wavelength of light. We further study the cross correlations
between two spatially distinct points that are generated due to 
propagation. It is found that such cross correlation
leads to sufficiently strong spectral dependence of polarization
which can be measured experimentally.
\end{abstract}

\bigskip
In a series of interesting papers Wolf\cite{Wolf86,Wolf87,Wolf87A} 
showed that in general the spectrum of
electromagnetic radiation does not remain invariant under propagation, 
even through vacuum. The effect arises if the source has spatial
correlations. The phenomenon was later confirmed experimentally
\cite{Faklis88,Gori88,Inde89} and has been a subject of considerable
interest \cite{review,Mandel}.
Further investigations of the source correlation
effects have been done in the time domain theoretically \cite{Rai}
and experimentally \cite{Chopra}.
Several applications of the effect have also
been proposed
\cite{James95,Kandpal95,Vicalvi96,Wolf97,Shirai98}. In a related
development it has been pointed out that spectral changes also
arise due to static scattering 
\cite{Foley89,Shirai95,Shirai96,Leskova97,Dogriu} and
dynamic scattering \cite{Wolf89,Foley89A,James90,JW90,JW94}.

In the current paper we study the spectral dependence of polarized
radiation that can arise due to spatial correlations.
We are interested in sources where the emission from different
points in the source are correlated i.e. the phase and amplitude 
shows systematic dependence on the position at the
source. This subject has attracted considerable attention
recently \cite{Jaiswal,James94,Gori20,Gori99,Gori98,Agrawal20,Gbur20}. In the
present paper we are interested in obtaining the form of the
correlation matrix for which the polarization in the far zone will 
not show any spectral dependence. 
One physical example we have in mind is radiation from a 
plasma in the presence of background magnetic field. 
The motion of charged particles at different spatial locations is 
in general correlated and will lead to spatially correlated
radiation.
In a separate paper \cite{Agarwal} we have also studied the
angular dependence of the polarization in the far zone due to
spatial correlations.

Consider a spatially extended 3-D source of polarized radiation, as shown in
figure 1, characterized by the charge density $\rho^{(r)}({\bf r},t)$ 
and current density ${\bf J}^{(r)}({\bf r},t)$. 
The electric and magnetic field vectors, ${\bf E}^{(r)}$ and ${\bf B}^{(r)}$, 
generated by this source, can be written as \cite{jackson},
\begin{eqnarray}
{\bf E}^{(r)}({\bf R},t) &=& {1\over 4\pi\epsilon_0}\int d^3 r 
\Bigg({{\bf S}\over S^3}
\left[\rho^{(r)}({\bf r},t')\right]_{\rm ret}
+{{\bf S}\over cS^2}\nonumber \\& &
\left[{\partial\rho^{(r)}({\bf r},t')\over \partial t'}\right]_{\rm ret} 
- {1\over c^2 S}
\left[{\partial{\bf J}^{(r)}({\bf r},t')\over \partial t'}\right]_{\rm ret}\Bigg)\\ 
{\bf B}^{(r)}({\bf R},t) &=& {\mu_0\over 4\pi}\int d^3 r
\Bigg(\left[{\bf J}^{(r)}({\bf r},t')\right]_{\rm ret}\times
{{\bf S}\over S^3}\nonumber\\
&+&\left[{\partial{\bf J}^{(r)}({\bf r},t')\over \partial t'}\right]_{\rm ret}
\times{{\bf S}\over c S^2}\Bigg)
\end{eqnarray}  
where ${\bf S} = {\bf R}-{\bf r}$, $S = |{\bf S}|$, 
$[f({\bf r},t')]_{\rm ret} = 
f({\bf r},t-S/c)$, $c=$ speed of light and ${\bf R}$ and 
${\bf r}$ are the position vectors of the observation point $Q$ (Fig 1) and a
particular point $P$ on the source respectively. 
Let ${\bf E}({\bf R},\omega)$,  ${\bf B}({\bf R},\omega)$, 
$\rho({\bf r},\omega)$ 
and ${\bf J}({\bf r},\omega)$ be the 
analytic signals associated with the Fourier transforms of the 
electric field, magnetic field, charge and current densities respectively.
In the radiation zone, ${\bf E}({\bf R},\omega)$ and  ${\bf B}({\bf R},\omega)$
are given by,
\begin{eqnarray}
{\bf E}({\bf R},\omega) &=& {i\omega\over 4\pi\epsilon_0c^2}\int d^3 r
{e^{i\omega S/c}\over R}\left[{\bf J}({\bf r},\omega)-{\bf J}({\bf r},\omega)
\cdot {\bf \hat R} {\bf \hat R}\right]
\label{Ew}\\
{\bf B}({\bf R},\omega) &=& {-i\omega\mu_0\over 4\pi c}\int d^3 r
{e^{i\omega S/c}\over R} {\bf J}({\bf r},\omega)\times {\bf \hat R}
\label{Bw}
\end{eqnarray}
where ${\bf \hat R}= {\bf R}/R$ and $R=|{\bf R}|$.
In obtaining this result we have used the current conservation 
equation and kept only the leading order terms in $1/R$.
The dimensions of the source (Fig. 1) are assumed to be much smaller than
its distance to the observation point $Q$ and hence dropping higher
orders in $1/R$ is reasonable. We see that the electric
and magnetic fields are orthogonal to one another as well as to
the propagation vector ${\bf k} = {\bf\hat R}\omega/c$.
Therefore at the far away point $Q$ 
only the $\theta,\phi$ components of the electric and magnetic field vector
are nonzero.

We are interested in evaluating the coherency matrix 
$J_{ij}({\bf R},\omega)$ at the point Q in the radiation zone. We can
relate this in terms of the source correlation function
\begin{equation}
W^S_{ij}({\bf r},{\bf r'};\omega) \delta(\omega-\omega')
= <J^*_i({\bf r},\omega)
J_j({\bf r'},\omega')> , 
\end{equation}
where the angular brackets represent ensemble averages.
Here we have made the standard assumption that 
the source fluctuations are represented by a stationary statistical ensemble, 
atleast in the wide sense \cite{Mandel}.
The correlation function of the electric field $W_{ij}$ 
is then also given by,
\begin{equation}
W_{ij}({\bf R},{\bf R'};\omega) \delta(\omega-\omega') = 
<E^*_i({\bf R},\omega)
E_j({\bf R'},\omega')> 
\end{equation}
The coherency matrix $J_{ij}({\bf R},\omega)$
 is given by,
\begin{equation}
{J}_{ij}({\bf R},\omega) = W_{ij}({\bf R},{\bf R},\omega)\ .
\end{equation}
A straightforward calculation 
using Eqs. \ref{Ew} and \ref{Bw} gives,
\begin{eqnarray}
{J}_{ij}({\bf R},\omega) &=& Z(\omega) 
\int d^3r d^3r' 
{e^{ik(S-S')}\over SS'}\nonumber\\
& &\xi_{il} W^S_{lm}({\bf r},{\bf r'},\omega)  \xi_{mj}
\label{Jij1}
\end{eqnarray}
where,
\begin{equation}
\xi_{ij} = -\hat R_i\hat R_j + \delta_{ij}\ ,
\end{equation}
$k=\omega/c$, $S'=|{\bf R}-{\bf r'}|$
and $Z(\omega)$ is an overall normalization factor which
will not play any role in our analysis. 
In Eq. \ref{Jij1} as well in the rest of the paper summation over 
repeated indices is understood unless otherwise stated.
We point out that since only
the transverse $(\theta,\phi)$ components of the electric field
vector are nonzero, the subscripts on $J$ 
as well as $W^S$ refer only to these
components. The
$(S-S')$ term in the exponent can be replaced by ${\bf\hat R}\cdot
{\bf \Delta}$,
where ${\bf\Delta}={\bf r'}-{\bf r}$.

The matrix $W^S_{ij}({\bf r},{\bf r'},\omega)$
measures the correlations between two spatially distinct points.
We are interested in studying the polarization at a point $Q$ in the far
zone when the cross correlation matrix has nontrivial dependence on
${\bf r}$ and ${\bf r'}$. 
It is convenient to express the matrix $W^S_{ij}$ in terms of the variables
${\bf r_a} = ({\bf r} + {\bf r'})/2$ and ${\bf \Delta}$ instead of 
${\bf r}$ and ${\bf r'}$. 
We rewrite Eq. \ref{Jij1} in terms of ${\bf r_a}$ and 
${\bf\Delta}$,
\begin{eqnarray}
{J}_{ij}({\bf R},\omega) &=& Z(\omega)
\int d^3r_a d^3\Delta 
{e^{ik{\bf\hat R}\cdot {\bf\Delta}}\over R^2}
\nonumber\\
& &\xi_{il} W^S_{lm}({\bf r_a},{\bf \Delta},\omega) \xi_{mj}\ .
\label{proof1}
\end{eqnarray}
The limits of integration over the variables ${\bf r_a}$ and ${\bf\Delta}$ 
are assumed to extend over all space. In obtaining Eq. \ref{proof1}
we have ignored higher order terms in $|{\bf \Delta}|/R$ in the exponent
which is reasonable if the observation point is far enough away.

In this paper we are primarily 
interested in studying the ${\bf \Delta}$ dependence of the correlation
matrix and hence consider only sources for which the ${\bf r_a}$
dependence factorizes, that is,
\begin{equation}
 W^S_{ij}({\bf r_a},{\bf \Delta},\omega) =  j_{il}({\bf r_a}) G_{lj}
({\bf \Delta},
\omega)\ . 
\label{factorized}
\end{equation}
This equation defines the two matrices,
$j_{ij}({\bf r_a})$ and $G_{ij}({\bf \Delta},
\omega)$. 
Physically this factorization means that the spatial correlations as well as
the spectral response of different regions of the source are identical to
one another. However the polarization of light emitted by different
points on the source need not be identical. 
Furthermore in order to focus on the spectral dependence
of polarization 
which arises due to spatial correlations we assume that, 
\begin{equation}
G_{ij}({\bf \Delta}=0 ,\omega) = N {\cal S}(\omega) \delta_{ij}
\label{Grr}
\end{equation}
where $N$ is a normalization factor, ${\cal S}(\omega)$ is the spectrum
of light emitted by any point on the source and $\delta_{ij}$ is the
standard Kronecker delta. Physically Eq. \ref{Grr}
means that if we ignore spatial correlations,  
the polarization of light emitted by 
the source has no spectral dependence. 
If the spatially distinct points are independent then,
\begin{equation}
 G_{ij}({\bf \Delta},\omega) = \delta^3({\bf\Delta}) 
{\cal S}(\omega)\delta_{ij}\ . 
\end{equation}
In this case we find that the resultant matrix $J$ is given
by
\begin{equation}
{J}_{ij}({\bf R},\omega) = {{\cal S}(\omega)\over R^2}\int d^3r_a
\xi_{il}j_{lm}({\bf r_a}) \xi_{mj}
\end{equation}
i.e. an incoherent integral over the entire source. 
  In this case we find, as expected, that the resulting 
polarization at $Q$ has no spectral dependence. 

\bigskip
\begin{figure}
\psfig{file=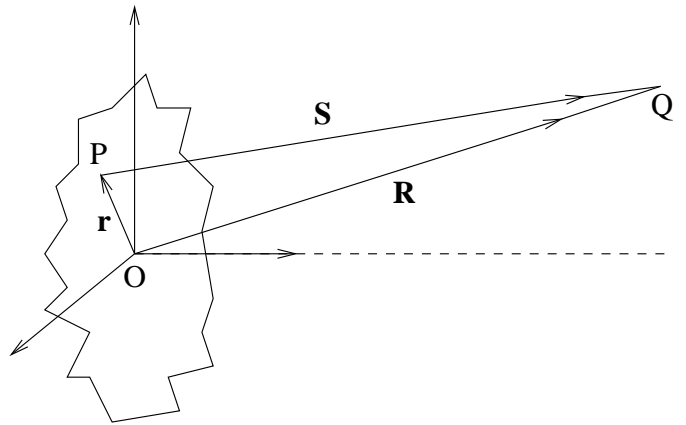}
\medskip
\caption{Schematic illustration of spatially extended 3-D source 
of polarized radiation and the observation point $Q$ located at 
position vector ${\bf R}$ from the source. $P$ represents any point on the
source located at position vector ${\bf r}$ with respect to the origin $O$ of
the coordinate system and at a distance $S=|{\bf S}|$ from Q.}
\end{figure}

In general, however, the polarization
observed at the far away point $Q$ does acquire spectral dependence
purely due to spatial correlations. We next obtain conditions
on the correlation matrix $W^S$ under
which this spectral polarization dependence is absent.  
We consider only those sources which satisfy the conditions
given in Eq. \ref{factorized} and \ref{Grr} 
since we are interested in isolating the
spectral dependence that arises only due to spatial correlations.

If the coherency matrix factorizes
into a function of $\omega$ times a matrix independent of $\omega$, that is,
\begin{equation}
{J}_{ij}({\bf R},\omega) = M_{ij}({\bf R}) h(\omega)\ ,
\label{JijFactorized}
\end{equation}
where $M_{ij}$ is a matrix independent of $\omega$ and $h(\omega)$
is a function of $\omega$,
then the polarization observed at $Q$ will have no spectral
dependence. 
In order to obtain the functional form of $W^S_{lm}({\bf r_a},{\bf \Delta},
\omega)$ which can lead to coherency matrix of the form given in 
Eq. \ref{JijFactorized} we substitute Eq. \ref{factorized} into 
Eq. \ref{proof1}. We can then express $J_{ij}$ as
\begin{eqnarray}
{J}_{ij}({\bf R},\omega) & =& Z(\omega) \xi_{il} {J^{0}_{ln} \tilde 
G_{nm}({\bf k},\omega)\over R^2} \xi_{mj}
\end{eqnarray}
where
\begin{equation}
J^{0}_{ln} = \int d^3r_a j_{ln}({\bf r_a})
\end{equation}
and 
\begin{equation}
\tilde G_{nm}({\bf k},\omega) = \int d^3\Delta G_{nm}({\bf \Delta},\omega)
e^{i{\bf k}\cdot \Delta},
\end{equation}
with ${\bf k} = k{\bf\hat R}$.
Hence we see that $\tilde G_{nm}({\bf k},\omega)$ is the Fourier 
transform of $G_{nm}({\bf \Delta},\omega)$. We can therefore also write
\begin{equation}
G_{nm}({\bf \Delta},\omega) = \int {d^3k\over (2\pi)^3} 
\tilde G_{nm}({\bf k},\omega) e^{-i{\bf k}\cdot {\bf \Delta}}
\label{Gnm}
\end{equation}
We point out that the integration in the above equation
is performed treating $\omega$ to be independent of ${\bf k}$.
In order that the polarization at the point $Q$ has no spectral 
dependence,
 $\tilde G_{nm}({\bf k},\omega)$ has to be of the form 
\begin{equation}
\tilde G_{nm}({\bf k},\omega) = \tilde A_{nm}({\bf k}/\omega) h(\omega)
\label{scaling}
\end{equation}
or
\begin{equation}
\tilde G_{nm}({\bf k},\omega) = \delta_{nm} \tilde H({\bf k},\omega) \ ,
\label{factorization}
\end{equation}
where the matrix $\tilde A_{nm}$ depends on ${\bf k}$ and
$\omega$ only through the combination ${\bf k}/\omega$ and $H({\bf k},\omega)$
is some function of ${\bf k},\omega$. Substituting Eq. \ref{scaling}
into Eq. \ref{Gnm} we find that,
\begin{equation}
G_{lm}({\bf \Delta},\omega) = h(\omega) 
G_{lm}(\omega{\bf\Delta}) \ \ ,
\label{scaling1}
\end{equation}
which is analogous to the scaling law obtained by Wolf \cite{Wolf86}
in his analysis of spectral shifts from spatially correlated
sources.  
Alternatively substituting the factorized form Eq. \ref{factorization}
into Eq. \ref{Gnm} we find that,
\begin{equation}
G_{lm}({\bf \Delta},\omega) = \delta_{lm} H({\bf \Delta},\omega) \ , 
\label{factorization1}
\end{equation}
where $H({\bf \Delta},\omega)$ is the Fourier transform
of $\tilde H({\bf k},\omega)$.
We therefore find that in order that the polarization in the 
radiation zone does not acquire spectral dependence, the correlation
matrix $G_{lm}({\bf \Delta},\omega)$ has to be of the form given
in Eq. \ref{scaling1} or Eq. \ref{factorization1}.

We next consider a specific example and calculate the spectral 
dependence arising due to correlations. We consider a planar circular source
of radius $a$,
which is spatially uncorrelated. The source emits polarized radiation
such that its coherency matrix
is given by,
\begin{equation}
  J(\rho,\phi) = A \left(\matrix{\sin^2\phi & -\sin\phi\cos\phi\cr
			-\sin\phi\cos\phi & \cos^2\phi}\right) \label{Jij3}
\end{equation}
where $A$ is a constant and $\rho,\phi$ are the polar coordinates
of any point at the position vector ${\bf r}$ on the source.
The source luminosity is independent of position and the polarization 
vectors point along ${\bf\hat\phi}$. 
We point out that the source has been constructed
such that at any point close to the axis of symmetry of the source 
the integrated polarization
is zero.

\bigskip
\begin{figure}
\psfig{file=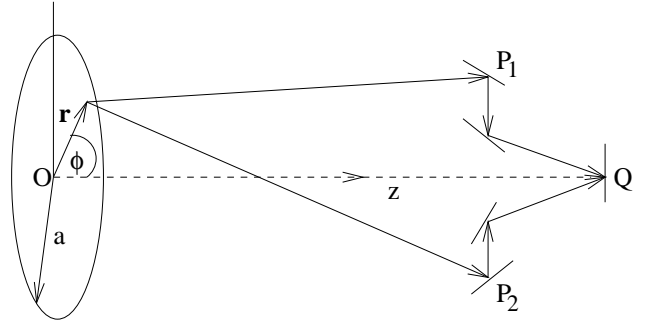}
\bigskip
\caption{Schematic illustration of a 2-D spatially extended
source of polarized light. $P_1({\bf R_1})$ and $P_2({\bf R_2})$ 
are any points close
to the axis of the source. After reflection from points $P_1$ and $P_2$,
light reaches the observation point $Q$, where its spectrum is measured. 
Due to spatial correlation between
$P_1$ and $P_2$, the polarization of light at $Q$ displays a nontrivial
spectral dependence.} 
\end{figure}

As is well known, although the source
is uncorrelated, the cross correlation between any two points $P_1$ 
and $P_2$ need not be zero due to the Van Cittert-Zernike theorem 
\cite{Mandel}. 
We consider the experimental arrangement 
shown in Fig. 2, where the light after being reflected from $P_1$ and
$P_2$ is observed at the point $Q$. We are interested in the spectral 
dependence of the polarization observed at $Q$.
The cross correlation matrix between any two points
$P_1({\bf R_1})$ and $P_2({\bf R_2})$ in the far zone close to symmetry axis
of the source, 
is given by,
\begin{equation}
W_{ij}({\bf R_1},{\bf R_2},\omega) = \left({k\over 2\pi}\right)^2
\int d^2{r} J_{ij}({\bf r}){e^{-ik (\rho/R) L
\cos(\phi-\psi)}\over R^2}\label{Wij}
\end{equation}
where 
$L\cos\psi=x_2-x_1$, $L\sin\psi=y_2-y_1$, $(x_1, y_1)$ and $(x_2,y_2)$
are the cartesian coordinates of the projections of ${\bf R_1}$ and ${\bf R_2}$
respectively on the plane of the source
and $R=R_1=R_2$ is the distance of the point $O$ on the 
source from points $P_1$ and $P_2$ which
have been assumed to be placed symmetrically for simplicity. 
In obtaining Eq. \ref{Wij} we have followed the treatment given
in Ref. \cite{Mandel} 
for the calculation
of cross correlation between two points $P_1$ and $P_2$ close
to the symmetry axis of a spatially uncorrelated source.
 Lack of spatial correlation implies that
the cross correlation matrix at the source $W^S_{ij}({\bf r_1},{\bf r_2},
\omega)=
J_{ij}({\bf r_1},\omega)\delta^2({\bf r_2}-{\bf r_1})$. We have further
assumed that $J_{ij}({\bf r_1},\omega)$ has no spectral dependence.
The integral in Eq. \ref{Wij} is over the source and 
we are using polar coordinates
$x = \rho\cos\phi$ and $y = \rho\sin\phi$. 

\bigskip
\begin{figure}
\psfig{file=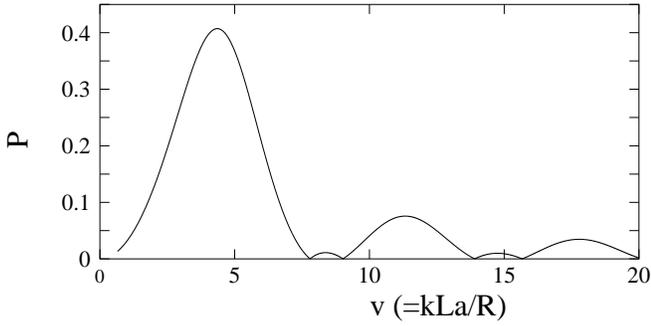}
\bigskip
\caption{The degree of polarization $P$ at the observation point $Q$ due to
the source model specified by Eq. \ref{Jij3} as a function of $v=kLa/R$.
Here $k$ is the wavenumber of light, L is the distance between points
$P_1$ and $P_2$ (see Fig. 2), $a$ is the radius of the primary
2-D source and $R$ is the distance of the points $P_1$ and $P_2$,
assumed to be located symmetrically, from
the center of the source. 
The 2-D source is spatially
uncorrelated with uniform intensity and polarization vectors 
pointing along $\hat \phi$ at any point $(\rho,\phi)$ on the source.}  
\end{figure}

Following the treatment given
in \cite{Mandel} we can calculate the cross correlation matrix and hence
the Stokes parameters at the point of observation. For the source
under consideration the result can be obtained analytically. We find,
upto an overall common factor $A a^2k^2/2\pi R^2$
$$s_0 = 1+ 2 J_1(v)/v\ ,$$ 
$$s_1={2\over v^2}{(y_2-y_1)^2 - (x_2-x_1)^2\over L^2}[vJ_1(v) - 2(1-J_0(v)]\ ,$$
$$s_2=-{4\over v^2}{(x_2-x_1)(y_2-y_1)\over L^2}[vJ_1(v) - 2(1-J_0(v)]\ ,$$
$$s_3 = 0$$
where $v= k L a/R$.
We therefore find that the wave is linearly polarized at the point of
observation $Q$ with the degree of polarization given by,
\begin{equation}
P = 2{|-vJ_1(v)+2-2J_0(v)|\over v^2+2vJ_1(v)}
\end{equation}
which has nontrivial spectral dependence. The
orientation of the linear polarization vector, given by,
\begin{equation}
\tan(2\psi) = {2(x_2-x_1)(y_2-y_1)\over (x_2-x_1)^2-(y_2-y_1)^2}\ ,
\end{equation}
 no spectral dependence.
The calculated degree of polarization for this example
is plotted in Fig. 3.
We clearly see that it is a very significant effect and can 
be observed experimentally. The orientation of the linearly polarized component
depends on the positions of $P_1$ and $P_2$ and contains information about
the polarization profile of the source.

\bigskip
\begin{figure}
\psfig{file=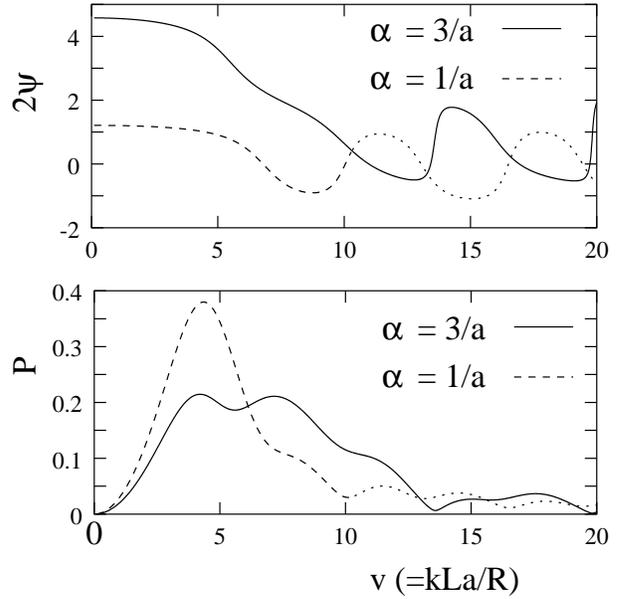}
\bigskip
\caption{(a) The linear polarization angle $\psi$ and 
(b) the degree of polarization $P$ at the observation point $Q$ 
as a function of $v=kLa/R$.
Here $k$ is the wavenumber of light, L is the distance between points
$P_1$ and $P_2$ (see Fig. 2), $a$ is the radius of the primary
2-D source and $R$ is the distance of the points $P_1$ and $P_2$,
assumed to be located symmetrically, from
the center of the source. 
The primary 2-D source is spatially
uncorrelated with uniform intensity, with its polarization
specified by Eq. \ref{Jij3} with $\phi$ replaced by
$\phi+\alpha\rho$, where $\alpha$ is a parameter. 
The polarization vectors 
make an angle $\alpha\rho$ with 
$\hat \phi$ at any point $(\rho,\phi)$ on the source. We study two 
representative choices of the parameter $\alpha=1/a,3/a$. 
}  
\end{figure}

We next study a somewhat more complicated
source for which the coherency matrix at any point ${\bf r}=(\rho,\phi)$ is
same as that given in Eq. \ref{Jij3}, with $\phi$ replaced by $\phi+\alpha \rho$,
where $\alpha$ is a parameter.
In this case we numerically calculate the Stokes parameter. The 
spectral dependence of the degree of polarization and the orientation
of the linear polarization is shown in Fig. 4 for some representative
choices of the parameter $\alpha$. 
The plot uses $x_2-x_1=1.0$ and $y_2-y_1=-0.2$ in 
arbitrary units. The state of polarization in the far zone 
only depends on the dimensionless
quantities $v=kLa/R$, $(x_2-x_1)/L$, $(y_2-y_1)/L$ and $\alpha a$
where $L^2 = (x_2-x_1)^2+(y_2-y_1)^2$.
We find that in this case both the
orientation angle of the linear polarization vector 
and the degree of polarization
show a dramatic spectral variation. 
The Stokes parameter $s_3$ vanishes in this case also
showing that there is no circularly polarized component at the point $Q$.

The effect discussed in this paper is observable experimentally
by using a primary source which has spatial correlations or by generating
the correlations due to propagation as shown in the above example. 
We assume an experimental arrangement shown in Fig. 2.
We first measure the spectral dependence of polarization at the point
$Q$ due to the waves emerging from the secondary sources $P_1$ 
and $P_2$ individually. The effect of spatial correlations can then
be determined by measuring the polarization at $Q$ due to 
interference of the waves emerging from $P_1$ and $P_2$. 
 
We conclude that spatially correlated sources of polarized radiation
generically display nontrivial spectral dependence  
of the state of polarization in the far zone. This dependence
goes away if the correlation
matrix displays a scaling law 
or factorizes into a constant matrix and a function of
the relative coordinate and frequency.

\end{document}